# Iron and molybdenum valences in double-perovskite $(Sr,Nd)_2FeMoO_6$: electron-doping effect

J. Lindén[1], T. Shimada[2], T. Motohashi[2], H. Yamauchi[2], and M. Karppinen[2],*

[1]*Physics Department, Åbo Akademi, FIN-20500 Turku, Finland*

[2]*Materials and Structures Laboratory, Tokyo Institute of Technology, Yokohama 226-8503, Japan*

Double perovskite, $(Sr_{1-x}Nd_x)_2FeMoO_6$, was doped with electrons through partial substitution of divalent Sr by trivalent Nd ($0 \leq x \leq 0.2$). The Fe valence and the degree of *B*-site order were probed by $^{57}$Fe Mössbauer spectroscopy. Replacing Sr by Nd increased the fraction of Fe and Mo atoms occupying wrong sites, *i.e.* antisite disorder. It had very little effect on the Fe valence: a small but visible increase in the isomer shift was seen for the mixed-valent $Fe^{II/III}$ atoms occupying the right site indicating a slight movement towards divalency of these atoms, which was more than counterbalanced by the increase in the fraction of antisite Fe atoms with III valence state. It is therefore argued that the bulk of the electron doping is received by antisite Mo atoms, which - being surrounded by six $Mo^{V/VI}$ atoms - prefer the lower IV/V valence state. Thus under Nd substitution, the charge-neutrality requirement inflicts a lattice disorder such that low-valent $Mo^{IV/V}$ can exist.

* Corresponding author:
Prof. Maarit Karppinen
Materials and Structures Laboratory
Tokyo Institute of Technology
4259 Nagatsuta, Midori-ku
Yokohama 226-8503, Japan
Phone: +81-45-924-5333
Fax: +81-45-924-5365
E-mail: karppinen@msl.titech.ac.jp

# I. INTRODUCTION

The halfmetallic properties and tunneling magnetoresistance (TMR) phenomenon of the *B*-site ordered double-perovskite, $Sr_2FeMoO_6$, and related $A_2FeMoO_6$ (*A*: alkaline earth element) compounds are well-established by now [1]. Initially an electronic structure based on localized $3d^5$ electrons of high-spin $Fe^{III}$ and an itinerant $4d^1$ electron of $Mo^V$ was assumed for $Sr_2FeMoO_6$. However, we showed that iron in $Sr_2FeMoO_6$ possesses a mixed-valence state, expressed as $Fe^{II/III}$, as seen by both $^{57}Fe$ Mössbauer [2] and Fe *K*- and *L*-edge XANES [3] spectroscopy probes. This view is now widely accepted [4,5]. The precise valence of the mixed-valent $Fe^{II/III}$ species is not necessarily fixed at 2.5, but it is rather controlled by the size of the *A* cation, *i.e.* employing "isovalent cation substitution". This was shown using a series of $(Sr,Ca,Ba)_2FeMoO_6$ samples [6,7]. Mössbauer data revealed another type of iron species too, *i.e.* trivalent Fe atoms or antisite atoms (AS) sitting at the lattice site reserved for Mo in the completely ordered double-perovskite structure [2,6,7]. Also obtained was from Mössbauer measurements experimental evidence for antiphase boundaries (APB) in $Sr_2FeMoO_6$ [8].

Efforts to enhance the TMR effect at room-temperature have typically been based on various substitution schemes, as well as methods for controlling the long-range ordering of the Fe and Mo atoms. Among the interesting concepts an electron-doping scheme of replacing Sr with a trivalent rare-earth element has been employed [9]. An increase in the ferrimagnetic ordering temperature ($T_C$) has been observed [9,10], despite the decrease in saturation magnetization. In the present work, $^{57}Fe$ Mössbauer spectroscopy has been applied to study the influence of electron doping, accomplished by partial replacement of Sr by Nd, on the valence states of Fe and Mo.

# II. EXPERIMENTAL

Samples of $(Sr_{1-x}Nd_x)_2FeMoO_6$ were synthesized by means of an oxygen-getter-controlled low-$O_2$-pressure encapsulation technique as developed earlier [11]. In brief, stoichiometric quantities of $SrCO_3$, $Nd_2O_3$, $Fe_2O_3$ and $MoO_3$ powders were mixed and calcined in air at 900 °C. The calcined powder mixture was pelletized and encapsulated in an evacuated fused-quartz ampoule with grains of Fe acting as a getter for oxygen, and fired at 1150 °C for 50 hours. The concentration of Nd was $x$ = 0.00, 0.05, 0.10 and 0.20. For the $x$ = 0.20 sample a somewhat higher synthesis temperature of 1200 °C was used. The phase purity and the degree of *B*-site order were checked by Rietveld refinement (using program RIETAN [12]) of the powder x-ray diffraction patterns. Magnetization measurements under applied fields of 0 to 5 T were performed at 5 K, using a SQUID magnetometer (Quantum Design: MPMS-XL). Curie temperatures were determined "thermogravimetrically" (Perkin Elmer: Pyris 1), *i.e.* by measuring the temperature at which an apparent weight change occurred when an external magnetic field was applied.

For the $^{57}$Fe Mössbauer measurements an absorber was made by spreading the powder, mixed with an epoxy resin, on an Al foil. The absorber thickness was ~20 mg/cm$^2$. Spectra were recorded in a transmission geometry using the maximum Doppler velocity of 11.15 mm/s. The absorber temperature was set at 77 K. A Cyclotron Compony $^{57}$Co:*Rh* (25mCi, Jan. 2002) source was used. The full Hamiltonian of combined electric and magnetic interactions was used to fit the spectra, with the total magnetic field experienced by the Fe nucleus (*B*), the chemical isomer shift relative to α-Fe at 300 K (δ), the quadrupole coupling constant (*eQV$_{zz}$*), the resonance line widths (Γ), and the relative intensities of the components (*I*) as fitting parameters. For each component a certain variation in the parameter *B* was allowed in order to simulate the fact that the internal field has a certain spread: a Gaussian distribution was assumed and its width (Δ*B*) was also introduced as a fitting parameter.

## III. RESULTS AND DISCUSSION

According to the x-ray diffraction patterns all samples were perfectly phase pure. Thus the solubility limit of Nd at the Sr site exceeded 20 %. However, for the highest Nd concentration a somewhat higher synthesis temperature was required to remove traces of impurity phases. According to the results of Rietveld refinement the long-range ordering of the Fe-Mo rock-salt structure was weakened upon increasing *x*. The long-range order parameter (*S*) is here defined by: $S = 2(\omega_{Fe} - 0.5)$, where $\omega_{Fe}$ is the occupancy of Fe at its right site. Thus, *S* = 0 when the Fe occupancy is 0.5 and *S* =1 when the occupancy is 1. In parallel with the decrease of *S* a decrease in the saturation magnetization is observed. Thus for *x* = 0, $M_s$ = 3.9 μ$_B$ was obtained at 5 K, whereas for *x* = 0.20 a decrease to $M_s$ = 2.7 μ$_B$ occurred. This decrease is well explained by assuming a ferrimagnetic interaction between the Fe and Mo atoms, and that the antisite atoms by being misplaced diminishes the saturation magnetization according to $M_s/\mu_B = 4 - 8(1-\omega_{Fe}) = 4\,S$ [6]. The $T_C$ values were found to increase as a function of *x*: 418 K (*x* = 0), 410 K (*x* = 0.05), 432 K (*x* = 0.10), and 453 K (*x* = 0.20). This observation is in accordance with the results of previous studies [9,10].

Mössbauer spectra of the pristine compound, $Sr_2FeMoO_6$, can be analyzed using four major spectral components [7]. The origin of the components is schematically illustrated in Fig. 1. Component M1 with an internal field of ~45 T and an isomer shift of ~0.70 mm/s is due to Fe atoms residing in the mixed-valence II/III state at the right site. Component AS with an internal field of more than 50 T and an isomer shift of 0.3-0.5 mm/s is assigned to trivalent Fe atoms occupying the wrong site, *i.e.* antisite. The portion of this component provides us with a measure of the degree of Mo-Fe disorder. Component M2 with an internal field of ~47 T and an isomer shift of 0.6 mm/s arises from the Fe atoms adjacent to AS Fe atoms. These atoms are considered to be in the mixed-valence II/III state. Component APB having an unusually low field, only ~2 T, and an isomer shift of 0.2-0.5 mm/s is considered to arise from trivalent Fe atoms at antiphase boundaries [8]. These four components were taken as a starting point for the analysis of the present spectra. In Fig. 2, the fitted 77-K spectra of the samples with $x = 0.05$ and 0.20 are shown. Large enhancement of Component AS upon increasing $x$ is immediately observed, in accordance with the loss of long-range order as seen in the Rietveld analysis result, and with the decrease in $M_S$ as seen in the magnetization data. In the raw data Component AS appeared with a large line broadening. However, best fitting was obtained when an additional satellite component, Component AS', was introduced, instead of letting the $\Delta B$ parameter of the single AS component acquire an abnormally large value. It does not change the interpretation of the final result whether one uses a single or two AS components. In Fig. 3, the overall concentration of the two antisite components, AS and AS', is given against the substitution level $x$. Also plotted is the degree of disorder, $1-S$, as obtained through Rietveld refinements. The large enhancement of the antisite components, AS and AS', signifies an equally large enhancement in the portion of Mo atoms located at their antisite, *i.e.* the Fe site.

In Table 1, the hyperfine parameters obtained from the fittings are presented. The isomer shift values of the main components, M1 and M2, are plotted in Fig. 4. A small increase is observed upon increasing $x$. It signifies a small movement of the $Fe^{II/III}$ state towards divalency. The increase is, however, very small as compared *e.g.* with the jump of ~0.12 mm/s in isomer shift value observed upon an isovalent substitution of Ba for Sr [7]. For iron atoms in the $Sr_2FeWO_6$ compound an isomer shift of ~1.0 mm/s was seen, confirming the $Fe^{II}$ state [13]. Through careful examination of the present spectra a weak ~22 T component with an isomer shift of ~1.0 mm/s was barely discerned. In fitting it to the data this component covered 1 to 2 % of the total spectral intensity. Nonetheless its presence remained uncertain, and it was therefore omitted. Generally, if it is assumed that only iron atoms are reduced, the shifts of components M1 and M2 towards divalency are not enough to account for the electron doping even though combined with the possible presence of a trace of a divalent Fe component. In fact the average Fe valence increases with increasing $x$ in $(Sr_{1-x}Nd_x)_2FeMoO_6$, as the concentration of trivalent AS Fe atoms increases.

Chemically an Mo atom at a high valence state of V/VI surrounded by six other $Mo^{V/VI}$ atoms as nearest cation neighbors is rather unfavorable. As the concentration of antisites is clearly increased by the Nd substitution, the charge-neutrality argument immediately tells us that the antisite Mo atoms should reside at a rather low valence state of presumably IV/V. In fact it should be less than V, otherwise an overall reduction would not occur. Although no direct measurement of the valence of Mo is presented here, we may easily illustrate the above reasoning using the intensities of the spectral components from Table 1 for calculating the average Mo valence. Components M1 and M2 reflect the concentration of $Fe^{II/III}$. The defect components, AS, AS' and APB, are due to Fe atoms with a valence state of III. In Fig. 5 plotted is the average Mo valence as a function of $x$ as obtained using the charge-neutrality requirement, the Mössbauer component intensities for the different Fe species, and the valence values as assumed above (*i.e.* 2.5 for M1 and M2, and 3 for AS, AS' and APB). The average Fe valence is also shown for comparison. It is reasonable to assume that the number of Mo atoms at defect (AS and/or APB) site(s) is equal to the number of defect Fe atoms. Using Fig. 5 and the component intensity data, the valence of the Mo atoms located at the defect site(s) is calculated as shown in Fig. 6, assuming the valence of the right-site Mo atoms to be 5.5. Up to $x = 0.1$ the values are reasonable. However, the valence value of 3.1 obtained for the $x = 0.20$ sample is clearly too low, but well accounted for by the fact that a slight electron doping on the $Fe^{II/III}$ and $Mo^{V/VI}$ species has been neglected, *i.e.* constant valence values of 2.5 and 5.5, respectively, were assumed for these atoms throughout the sample series, $0 \leq x \leq 0.2$. If one assumes that *e.g.* 25 % of the doped electrons enter the mixed-valence species the valence of the defect Mo atoms would approach IV. The slight increase in the isomer shift of Component M1 indicates that this most probably is the case. Also the fact that $T_C$ increases slightly as a function of $x$ indicates that some of the doped electrons are delocalized in the halfmetallic conduction band, thereby strengthening the overall double-exchange mechanism. One should also note that component APB covers more than 10 % in the Mössbauer spectrum for $x = 0.10$ and only 5 % for $x = 0.20$, and although the concentration of AS and AS' atoms has increased from $x = 0.10$ to $x = 0.20$ there is a stronger pressure on the $Fe^{II/III}$ atoms at $x = 0.20$ to decrease their valence state, as there are relatively few antisite Mo atoms to carry the heavy electron doping. The concentration of APB atoms is not very easily controlled: the presence of APB is confirmed in both substituted and nonsubstituted samples.

## IV. CONCLUSIONS

Phase-pure samples of the electron-doped $(Sr_{1-x}Nd_x)_2FeMoO_6$ system were synthesized by partially substituting $Nd^{III}$ for $Sr^{II}$. Only a slight reduction in the valence value of the Fe atoms at the right site towards II was caused by this kind of electron doping. On the other hand, electron doping was seen to decrease the degree of Fe-Mo order markedly. The large reduction in the degree of order was interpreted as a result of strive for increasing the number of low-valent antisite $Mo^{IV/V}$ atoms that are the species suspected to receive the bulk of electron doping. Simultaneous decrease in the saturation magnetization supports this picture.

## ACKNOWLEDGMENTS

This work was supported by Grants-in-aid for Scientific Research (Nos. 15206002 and 15206071) from Japan Society for the Promotion of Science. J. L. acknowledges support from the Scandinavia-Sasakawa foundation.

**Table 1.** Hyperfine parameters obtained for the $(Sr_{1-x}Nd_x)_2FeMoO_6$ samples from the computer fittings of $^{57}Fe$ Mössbauer spectra at 77 K.

|  |  | Substitution level x |  |  |  |
|---|---|---|---|---|---|
| Comp. |  | 0 | 0.05 | 0.10 | 0.20 |
| M1 | $B$ (T) | 45.24(3) | 45.8(3) | 45.41(3) | 44.97(1) |
|    | $\delta$ (mm/s) | 0.700(6) | 0.712(3) | 0.706(4) | 0.733(1) |
|    | $I$ (%) | 77(4) | 75(2) | 57(3) | 47(1) |
|    | $eQV_{zz}$ (mm/s) | -1.0(2) | -0.2(2) | -0.3(2) | -0.6(1) |
| M2 | $B$ (T) | 45.6(3) | 49.0(2) | 48.0(2) | 47.9(8) |
|    | $\delta$ (mm/s) | 0.50(1) | 0.66(2) | 0.58(3) | 0.60(1) |
|    | $I$ (%) | 11(2) | 10(1) | 16(1) | 32(5) |
|    | $eQV_{zz}$ (mm/s) | 0.0(1) | 0.4(2) | -0.1(2) | 0.1(8) |
| AS | $B$ (T) | 51.7(2) | 52.6(5) | 51.0(x) | 51.7(2) |
|    | $\delta$ (mm/s) | 0.4(4) | 0.5(6) | 0.4(1) | 0.55(3) |
|    | $I$ (%) | 2.3(4) | 4.4(6) | 5.8(9) | 4.9(4) |
|    | $eQV_{zz}$ (mm/s) | -1(2) | -0.3(2) | -0.2(4) | 0.2(1) |
| AS' | $B$ (T) | - | 54.9(1) | 53.9(6) | 53.8(1) |
|     | $\delta$ (mm/s) | - | 0.47(2) | 0.47(5) | 0.52(2) |
|     | $I$ (%) | - | 3.9(5) | 6.1(6) | 8.8(3) |
|     | $eQV_{zz}$ (mm/s) | - | -0.2(9) | -0.1(2) | 0.1(1) |
| APB | $B$ (T) | 1.2(3) | 2.5(1) | 1.8(6) | 0.9(1) |
|     | $\delta$ (mm/s) | 0.34(4) | 0.54(2) | 0.38(5) | 0.26(2) |
|     | $I$ (%) | 9(1) | 5.1(5) | 12.9(6) | 5.1(3) |
|     | $eQV_{zz}$ (mm/s) | -0.8(1) | 1.0(9) | 1.5(2) | 0.8(1) |

**Figure captions**

**Fig. 1.** Schematic illustration of the origin of the various Fe species distinguished by Mössbauer spectroscopy for the $A_2$FeMoO$_6$-phase samples. The filled/empty spheres are for Fe/Mo atoms, O atoms are omitted.

**Fig. 2.** Mössbauer spectra recorded at 77 K for the $(Sr_{1-x}Nd_x)_2$FeMoO$_6$ samples with $x$ = 0.05 and 0.20. The components used in the fittings are displayed above each spectrum and labelled in the legend. The components and the labels are presented in the same order, *i.e.* the topmost is AS', followed by AS, *etc*.

**Fig. 3.** Total intensity of the two antisite components, AS and AS', (filled circles) and the degree of disorder, 1-$S$, (open squares) for the $(Sr_{1-x}Nd_x)_2$FeMoO$_6$ samples with increasing Nd content, $x$.

**Fig. 4.** Isomer shift values of the two main mixed-valence components, M1 (filled circles) and M2 (open circles), as obtained from the fitted Mössbauer spectra for the $(Sr_{1-x}Nd_x)_2$FeMoO$_6$ samples with increasing Nd content, $x$.

**Fig. 5.** Average valences of Fe (open squares) and Mo (filled circles) as estimated from the Mössbauer component intensities (requiring charge neutrality) for the $(Sr_{1-x}Nd_x)_2$FeMoO$_6$ samples with increasing Nd content, $x$.

**Fig. 6.** Valence value of the defect Mo atoms as estimated for the $(Sr_{1-x}Nd_x)_2$FeMoO$_6$ samples based on the Mössbauer data for the portion and valence of different Fe species; for the details see the text.

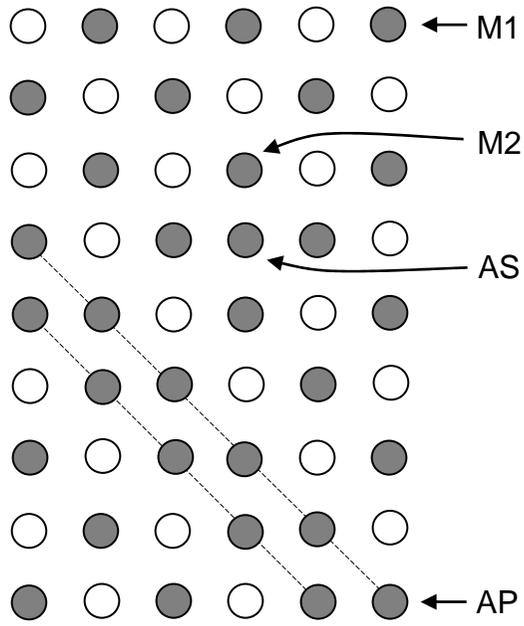

Fig. 1, Linden et al.

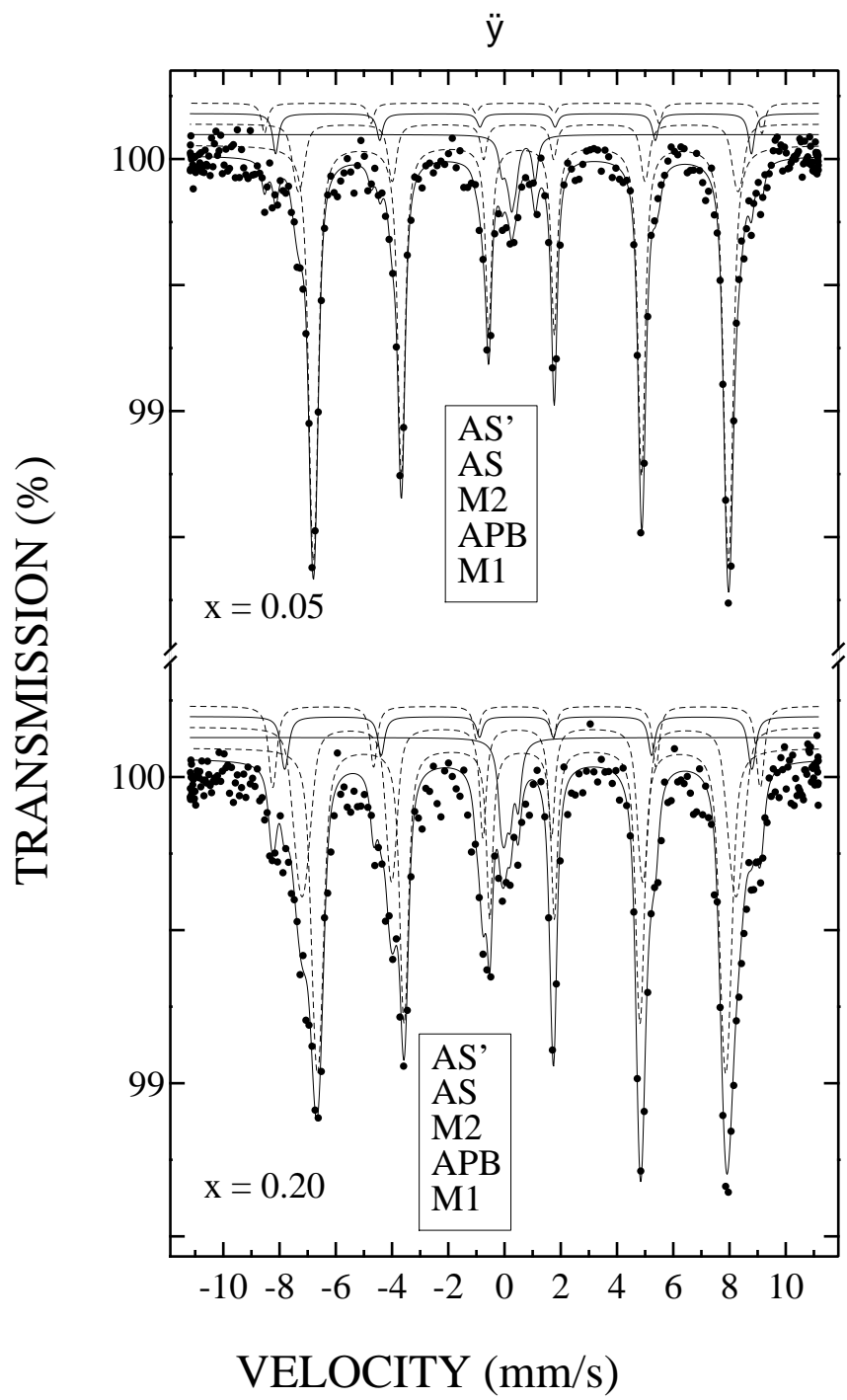

Fig. 2, Linden et al.

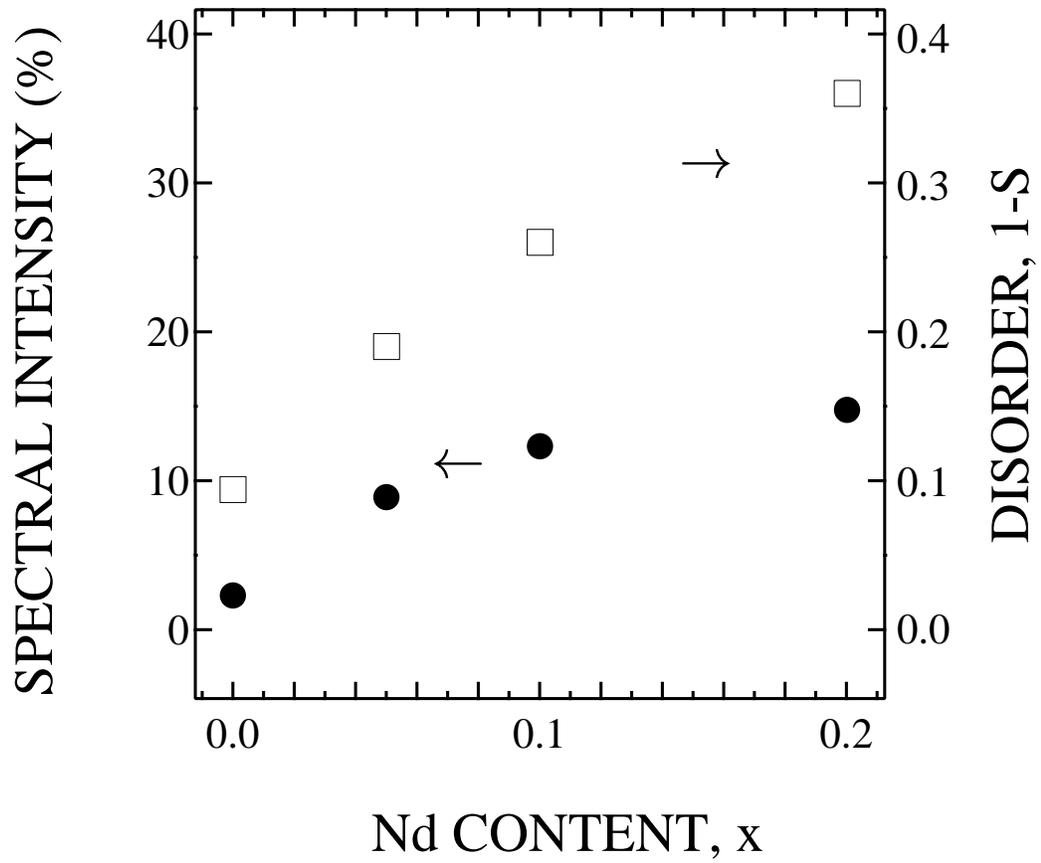

Fig. 3, Linden et al.

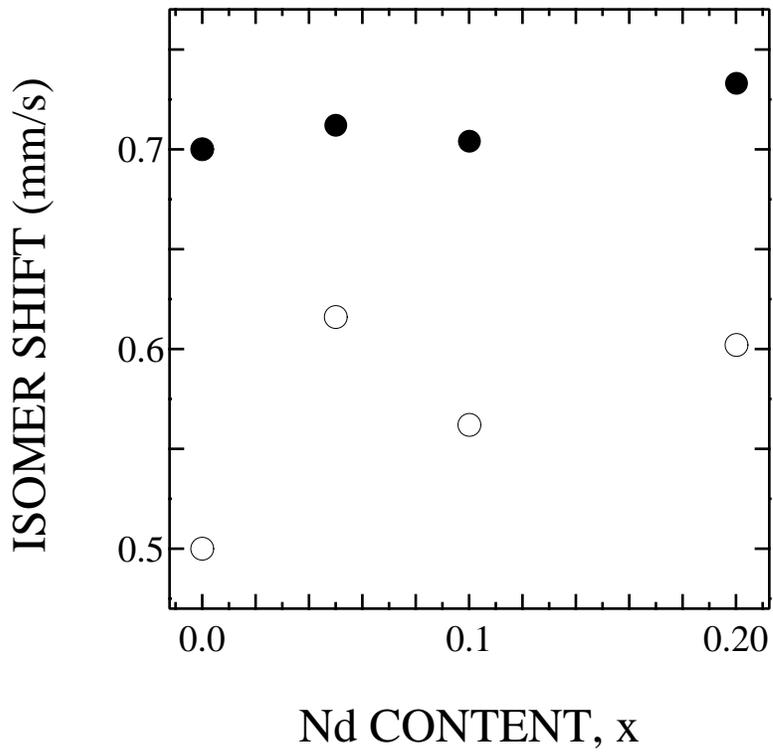

Fig. 4, Linden et al.

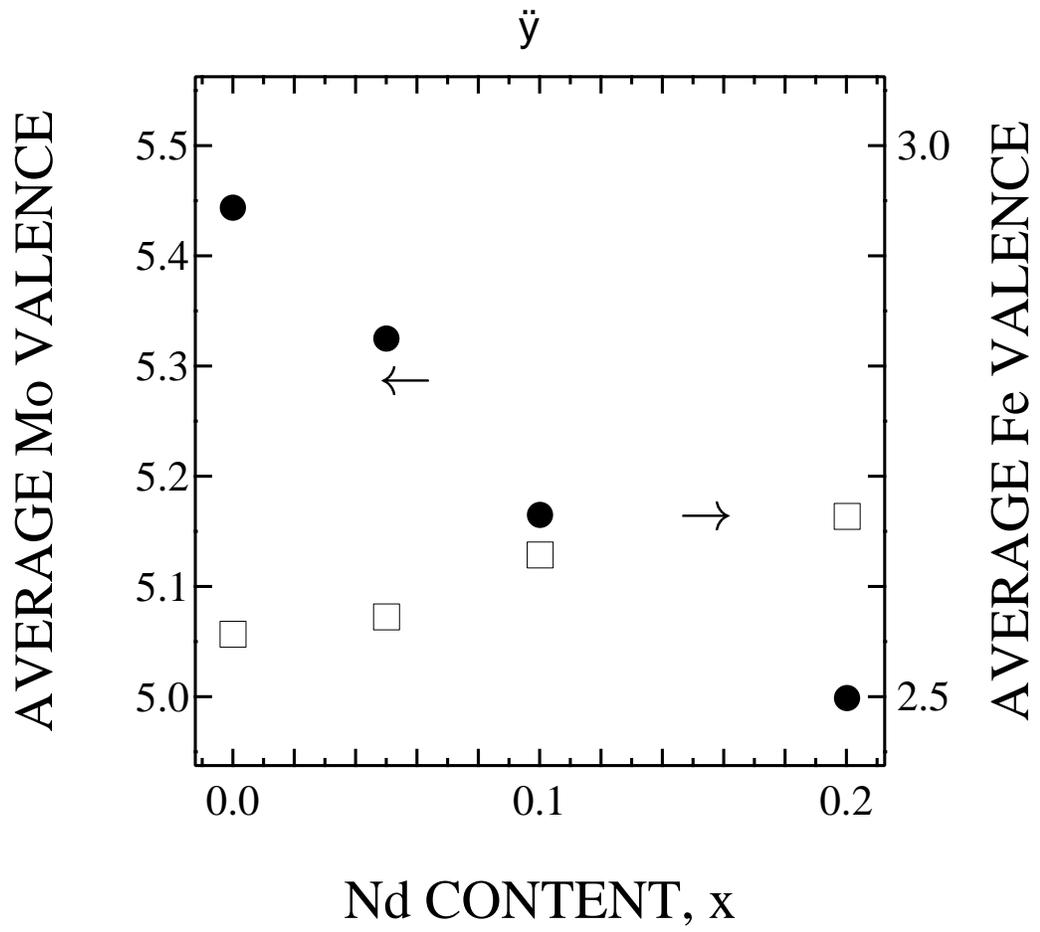

Fig. 5, Linden et al.

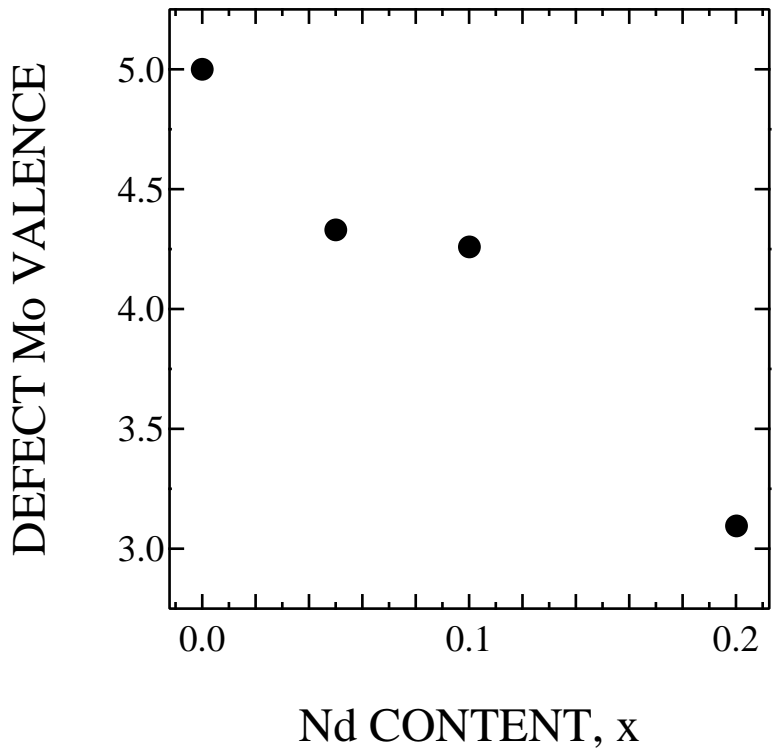

Fig. 6, Linden et al.